\documentclass[aps,prl,reprint,groupedaddress]{revtex4-1}
\usepackage{amssymb}
\usepackage{amsmath}
\usepackage{epsfig} 
\usepackage{epic,eepic}

\usepackage{hyperref}
\usepackage{lipsum}
\usepackage{fancyhdr}
\usepackage[utf8]{inputenc}
\usepackage[english, vietnamese]{babel}

\begin{document}


\title{Fermionic Lensing in Smooth Graphene P-N Junctions}


\author{Võ Tiến Phong}
\author{Jian Feng Kong}
\email[Corresponding author: ]{jfkong@mit.edu}
\affiliation{Massachusetts Institute of Technology, Department of Physics, Cambridge, Massachusetts 02139}


\date{October 1, 2016}

\begin{abstract}
Focusing of electron waves in graphene p-n junctions is a striking manifestation of fermionic negative refraction. We analyze lensing in smooth p-n junctions and find that it differs in several interesting ways from that in the previously studied sharp p-n junctions. Most importantly, while the overall negative-refraction behavior remains unchanged, the image at the focal point undergoes additional broadening due to Klein tunneling in the junction. We develop a theory of image broadening and estimate the effect for practically interesting system parameter values.
\end{abstract}


\maketitle


\begin{otherlanguage}{english}

Focusing of ballistic electrons has been extensively explored using electrostatic and magnetostatic lensing in two-dimensional electron gas (2DEG) systems \cite{PhysRevB.39.8556, content/aip/journal/apl/56/13/10.1063/1.102538, PhysRevB.70.041301, PhysRevB.92.235416}. Unfortunately, this technique faces some practical challenges from low operating temperatures, high fields, and sophisticated equipment. Using semiconductor p-n junctions (PNJs) as electron lens provides some solutions but has other problems: the large band gap and wide depletion region of conventional semiconductors are  too ``opaque” for effective electron focusing \cite{rudan2014physics}. Due to its unique properties, graphene has been proposed to overcome these shortcomings as an electronic lens \citep{Cheianov1252}. One such property is Klein tunneling, where a graphene PNJ is transparent to ballistic electrons, even for a smooth junction \cite{PhysRevB.74.041403, katsnelson2006chiral, 2053-1583-1-1-011006, :/content/aip/journal/jap/118/15/10.1063/1.4933395}. Recent experimental efforts have successfully demonstrated the use of graphene as an electronic lens \cite{lee2015observation, Chen1522}.

Theoretical efforts have been focused on sharp graphene PNJs \citep{Cheianov1252, :/content/aip/journal/jap/118/15/10.1063/1.4933395}, however, in realistic experimental set-up, the PNJ interface width is always finite. For instance, PNJs experimentally fabricated using suspended graphene have interface width on the order of 200 nm. With boron nitride (BN) encapsulation, a finer interface width can be achieved $\sim$ 20 nm \cite{lee2015observation}. An understanding of smooth PNJs is thus necessary to fully realize fermionic focusing in practical applications.

In this paper, we demonstrate that a smooth PNJ exhibits negative refraction and lensing similar to a sharp PNJ, as illustrated schematically in Fig.~\ref{fig: focusing in smooth junction}. However, the effect of junction smoothness is to blur the image at the focal point. In particular, for an incident wavepacket of size $a_0$, its image is broadened to a width $a$ given by
\begin{equation}\label{eq: image size}
a^2 = a_0^2 + \frac{\pi v_F \hbar}{eE},
\end{equation}
where $v_F \approx 10^6$ m/s, and $E$ is the electric field strength near the interface ($eE >0$). Our result illustrates that electron focusing in a smooth PNJ can be achieved even in the absence of junction curvature.

The result in Eq.~\ref{eq: image size} can be understood using a simple heuristic argument. We consider a smooth graphene PNJ that can be linearly approximated by constant electric field $E$, $U(x) = -eEx,$ near the junction. We assume that $U(x)$ is asymptotically constant at large $x$. Let us start with a wave packet with size $a$ incident from the left side of the junction $\psi(y) \propto e^{-y^2/2a_0^2}.$ In Fourier space, the components are $\psi(k_y) \propto e^{-k_y^2a_0^2/2}.$ Each harmonic after tunneling through the junction acquires a transmission factor \cite{schwinger1951gauge, 1969JETP...30..660N}
\begin{equation}\label{eq: transmission}
t(k_y) = \exp \left( - \frac{\pi v_F \hbar k_y^2 }{2eE} \right).
\end{equation} 
This formula, which is exact for a linear  potential, remains valid for a wide range  of $E$ and $k_y$ when the potential is smooth. The harmonics on the other side of the junction then become $\psi(k_y) \propto t(k_y) e^{-k_y^2a_0^2/2} =  e^{-k_y^2a^2/2},$ where $a$  given by Eq.~\ref{eq: image size}. Fourier transforming to position space, we see that $a$ is indeed the size of the image. This argument shows that the image is broadened after tunneling by a factor inversely proportional to the field strength $E.$ We will derive this result carefully, and illustrate that electron focusing in graphene is a robust phenomenon.

\begin{figure}
\begin{center}
\includegraphics[scale=0.567]{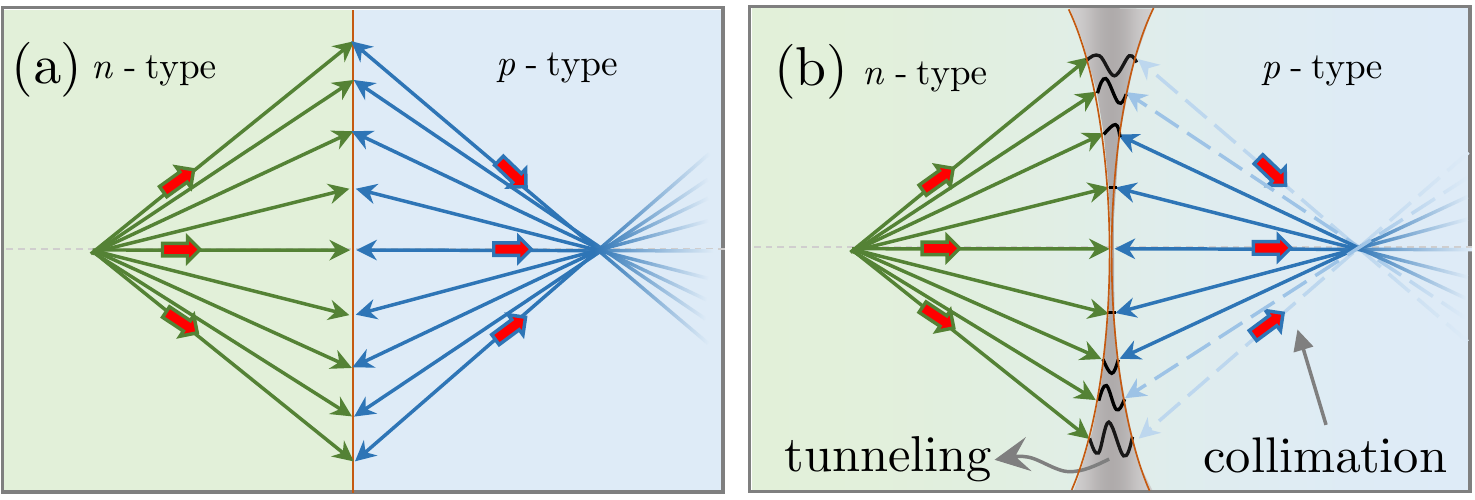}
\end{center}
\caption{Comparison of focusing in sharp and smooth graphene PNJs. (a)-(b) show focusing in a PNJ, where the wavevectors in the conduction band $\mathbf{k}_c$ and the wavevectors in the valence band $\mathbf{k}_v$ are on the $n$-type and $p$-type side respectively. In this case, both a sharp PNJ (a) and a smooth PNJ (b) produce focusing at the focal point. However, in the smooth junction, waves coming from the left undergo tunneling near the junction, whereby the transmission is suppressed exponentially by a factor given in Eq.~\ref{eq: transmission}. This is illustrated in (b) by the blurring out of wave components incident at large angles. }
\label{fig: focusing in smooth junction}
\end{figure}

In graphene's band structure, at the corners of its Brillouin zone, the conduction band and valence band are degenerate \cite{PhysRev.71.622, PhysRevB.46.1804}, as shown in Fig.~\ref{fig: Dirac cone}a. In undoped graphene, the Fermi level crosses these degeneracy points, and is shared by both bands. Fermions in this energy regime behave as relativistic Dirac particles obeying the Hamiltonian \cite{RevModPhys.81.109}, $\mathcal{H} = v_F \boldsymbol{\sigma} \cdot \mathbf{p},$ 
where $v_F \approx 10^6$ m/s, $\boldsymbol{\sigma} = (\sigma_x, \sigma_y)$ are the Pauli matrices in pseudospin (sublattice) space, and $\mathbf{p} = (p_x, p_y)$ are the momentum operators. The energy dispersion for electrons with quasi-momentum $\hbar \mathbf{k}= \hbar (k_x, k_y)$ in the conduction band is $\varepsilon_c = v_F \hbar |\mathbf{k}|;$ likewise, the dispersion for holes in the valence band is $\varepsilon_v = -v_F \hbar |\mathbf{k}|.$ Carrier densities in graphene can be precisely tuned through electrical gating \cite{novoselov2005two} or doping \cite{ ohta2006controlling, williams2007quantum}. 

 In the presence of an electrostatic potential energy $U(x)$ produced by electrical gating, the degeneracy points are shifted as shown in Fig.~\ref{fig: Dirac cone}a. The region where $U(x) < \mathcal{E},$ with $\mathcal{E}$ being the chemical potential, is of $n$-type with an electron density $\rho_c \sim k_c^2,$ where $k_c$ is the Fermi radius of the conduction band far away from the junction. Likewise, the $p$-type region is where $U(x) > \mathcal{E},$ and the hole density is $\rho_v \sim k_v^2.$ Suppose we have a sharp PNJ. An electron wave on the $n$-type side with wave vector $\mathbf{k} = k_c (\cos\theta_i, \sin \theta_i)$ has corresponding wave vector $\mathbf{k} = -k_v (\cos\theta_t, \sin \theta_t)$ on the other side of the junction. Due to symmetry in the $y$-direction, the component of $\mathbf{k}$ parallel to the junction is conserved, leading to the fermionic analogue of Snell's law for refraction
\begin{equation}
\label{Snell's law}
\frac{\sin \theta_i}{\sin \theta_t} = - \frac{k_v}{k_c} = n.
\end{equation}
As shown in Eq.~\ref{Snell's law}, the refractive index $n$ is negative, indicating that graphene can be used as a Veselago lens \cite{veselago1968electrodynamics} to focus electron flows, as shown in Fig.~\ref{fig: focusing in smooth junction}.

\begin{figure}
\begin{center}
\includegraphics[scale=0.567]{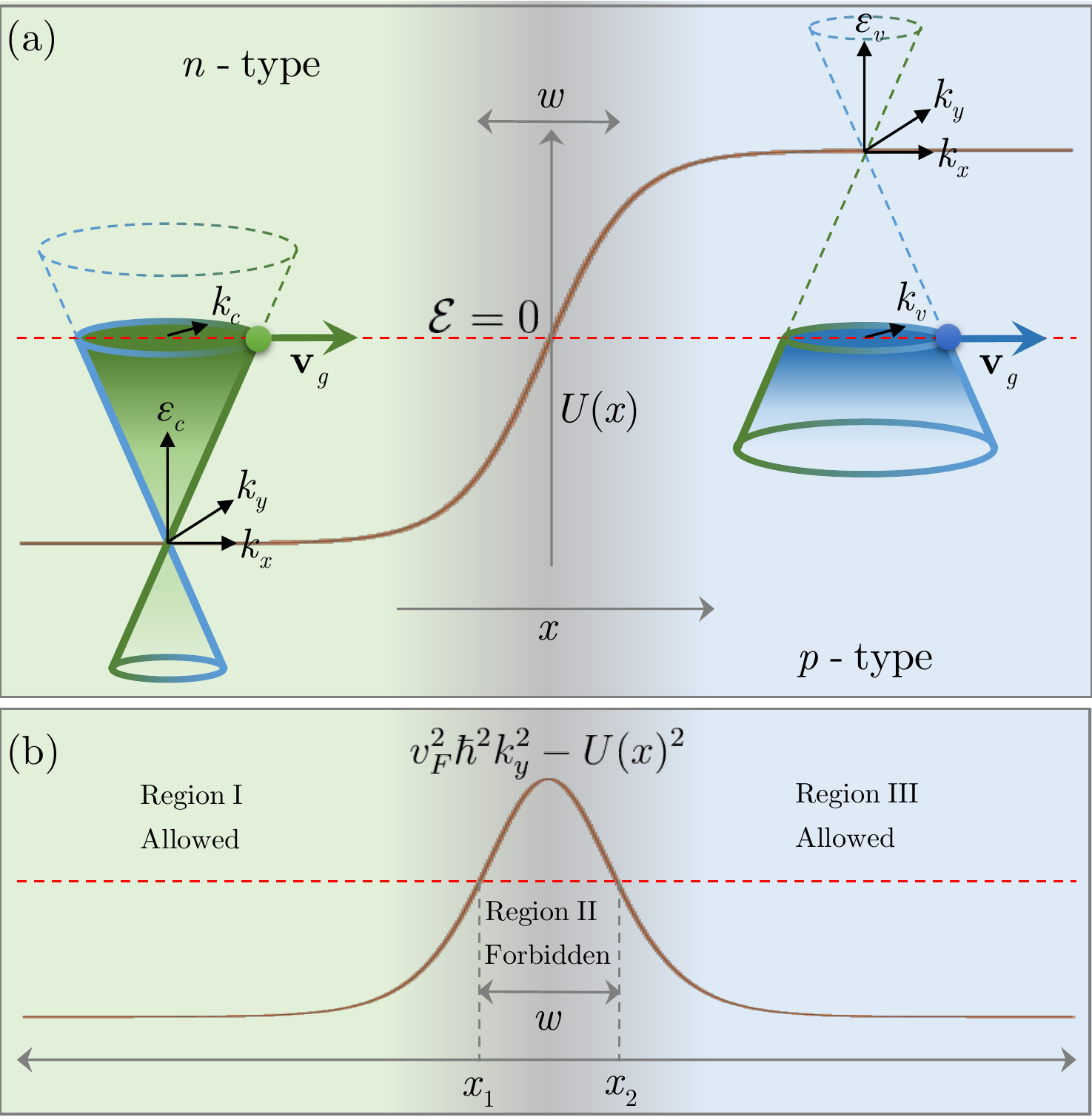}
\end{center}
\caption{(a) Dirac points are shifted by an electrostatic potential energy $U(x)$ produced by gating. The chemical potential is $\mathcal{E} = 0.$  (b) This illustrates the allowed regions of Dirac fermions when $p(x)^2 > 0,$ with $x_1$ and $x_2$ as the classical turning points.  }
\label{fig: Dirac cone}
\end{figure}

In optics, a Veselago lens is superior to a conventional lens because it can achieve a resolution below the diffraction limit \cite{pendry2000negative}. This is possible because evanescent Fourier modes are amplified by a Veselago lens so that when an image is reconstructed on the other side of the lens, all of the Fourier modes of the original field contribute, not just those of propagating waves. A Veselago lens is thus a so-called ``superlens" or perfect lens, because sub-wavelength imaging is possible, and the size of the image, $a,$ is equal to the size of the source, $a_0,$ i.e. $a = a_0$ and $a \ll \lambda$, where $\lambda$ is the wavelength of the imaging light. A similar story can be told for sharp graphene PNJs, except that fermionic waves are being focused instead of electromagnetic waves.

If the graphene PNJ is smooth, then the image resolution is reduced: $a>a_0.$ This is due to tunneling of fermions near the junction, and we have already heuristically  established this fact above. Since our argument relies on Eq.~\ref{eq: transmission}, we will now show a semi-classical derivation of it in the paraxial approximation. We consider the case where the junction is smooth, with width $w$ larger than the Fermi wavelength $\lambda_F.$ The opposite limit, $\lambda_F \gg w,$ corresponds to the sharp junction limit. Since $\lambda_F \sim \rho^{-1/2},$ where the fermion density $\rho$ can be easily tuned using gating \cite{RevModPhys.83.407}, $\lambda_F$ can always be tuned to be smaller than $w$ experimentally. In this case, we will show that wave vectors which are not perpendicular to the junction undergo quantum tunneling near the junction. However, a diverging wave-front on one side still converges on the other side, resulting in focusing of fermions shown in Fig.~\ref{fig: focusing in smooth junction}b. The quality of the image at the focal point depends on the sharpness of the junction, $|\partial_x U(x)|.$

Charge carriers in graphene are described by the Dirac equation
\begin{equation}
\label{eq: Dirac}
[v_F \boldsymbol{\sigma} \cdot \mathbf{p} +U(x)] \psi(x,y) = \mathcal{E} \psi(x,y),
\end{equation}
where without loss of generality, we can set $\mathcal{E} = 0.$ Analogous to methods in Fourier optics \cite{goodman2005introduction}, it is possible to obtain approximate solutions that illustrate focusing properties. To obtain semi-classical solutions, we first multiply Eq.~\ref{eq: Dirac} by $[v_F \boldsymbol{\sigma} \cdot \mathbf{p} -U(x)]$ to write 
\begin{equation}
\label{eq: Dirac2}
\left[ v_F^2 \mathbf{p}^2 - U(x)^2 - iv_F \hbar \frac{ dU}{dx} \right] \psi(x,y) = 0.
\end{equation}
Clearly, this transformation is not unitary. We have arrived at an effective Schrodinger equation that is non-Hermitian. However, as shown in \cite{1402-4896-2012-T146-014010}, this transformation is still useful in studying chiral tunneling in single-layer graphene. Since the Hamiltonian has translational invariance in the $y-$direction, we can Fourier transform $\psi$ in the $y-$direction
\begin{equation}
\label{eq: transform}
\psi(x,y) = \int \frac{dk_y}{2 \pi} e^{i k_y y} \psi (x, k_y).
\end{equation}
Substituting Eq.~\ref{eq: transform} into Eq.~\ref{eq: Dirac2}, we find the fermionic analogue of the Helmholtz equation \citep{goodman2005introduction} in optics
\begin{equation}
\label{eq: Helmhotz}
\frac{d^2 \psi(x,k_y)}{dx^2} = - \frac{1}{\hbar^2} \left( p(x)^2 + i \sigma_x \frac{\hbar}{v_F} \frac{dU}{dx} \right) \psi(x,k_y),
\end{equation}
where $p(x)^2 = [U(x)/v_F]^2 - \hbar^2 k_y^2,$ the classical momentum in the $x$-direction. We now diagonalize Eq.~\ref{eq: Helmhotz} by writing
\begin{equation}
\label{eq: diagonalization}
\psi(x,k_y) = \left( \begin{array}{ccc}
1  \\
1  \end{array} \right) \eta_1(x,k_y)+ \left( \begin{array}{ccc}
1  \\
-1  \end{array} \right) \eta_2 (x,k_y).
\end{equation}
The matrix equation in Eq.~\ref{eq: Helmhotz} now becomes scalar equations for $\eta_1$ and $\eta_2$
\begin{equation}
\label{eq: Helmhotz for eta}
\frac{d^2 \eta_{1,2}(x,k_y)}{dx^2} = - \frac{1}{\hbar^2} \left( p(x)^2 \pm i  \frac{\hbar}{v_F} \frac{dU}{dx} \right) \eta_{1,2}(x,k_y).
\end{equation}
To ensure self-consistency, we substitute Eq.~\ref{eq: diagonalization} into Eq.~\ref{eq: Dirac} to obtain the relationship between $\eta_1$ and $\eta_2$
\begin{equation}
\label{eq: self-consistency}
\eta_2(x,k_y) = \frac{1}{k_y} \left( \frac{d}{dx} + \frac{i}{\hbar k_y} U(x) \right) \eta_1(x,k_y).
\end{equation}
Eq.~\ref{eq: Helmhotz for eta} can be solved approximately using the Jeffreys-Wentzel–Kramers–Brillouin (JWKB) method with care taken when selecting appropriate propagating modes for hole and electron regions. We use the ansatz $\eta_1(x, k_y) = \exp [i S(x, k_y)/\hbar]$ and expand the phase in a power series of $\hbar,$ $S(x,k_y) = \sum_{j=0}^\infty \hbar^j S_j(x,k_y).$ In the allowed regions where $v_F\hbar |k_y| < |U(x)|,$ we have traveling waves of electrons (if $U(x) < 0$) and holes (if $U(x) >0$). When $v_F\hbar |k_y| > |U(x)|,$ no plane-wave solutions are allowed, and we instead have a decaying wavefunction. The electron wavefunction on the left side of the PNJ with boundary condition at $x = x_0$ is
\begin{equation}
\label{eq: wavefunction in allowed region}
\eta_1 \sim \frac{1}{\sqrt{p(x)}} \exp \left( \frac{i}{\hbar}\int_{x_0}^{x_1} \phi(\xi) d\xi +\frac{i}{\hbar}\int_{x_1}^{x} \phi(\xi) d\xi \right),
\end{equation}
where
\begin{equation}
\phi(x) = p(x) - \frac{\hbar}{2v_Fi} \frac{\partial_xU}{p(x)}.
\end{equation}
The corresponding forward-propagating hole wavefunction on the right side of the PNJ is 
\begin{equation}
\label{eq: wavefunction in forbidden region}
\eta_1 \sim \frac{t}{\sqrt{p(x)}} \exp \left( \frac{i}{\hbar}\int_{x_0}^{x_1} \phi(\xi) d\xi - \frac{i}{\hbar}\int_{x_2}^{x} \phi(\xi) d\xi \right),
\end{equation}
where 
\begin{equation}
t \sim \exp\left( \frac{i}{2v_F}\int_{x_1}^{x_2} \frac{\partial_x U}{|p(\xi)|} d\xi \right) \exp\left( - \frac{1}{\hbar}\int_{x_1}^{x_2} |p(\xi)| d\xi \right)
\end{equation}
is the transmission amplitude. The spinor wavefunction can be found using Eqs.~\ref{eq: diagonalization}, ~\ref{eq: self-consistency}, ~\ref{eq: wavefunction in allowed region}, and~\ref{eq: wavefunction in forbidden region}.

To study lensing from a PNJ, we choose a model potential $U(x)$ such that $U \rightarrow U_\infty$ for large $x$ and $U \rightarrow U_{-\infty}$ for large $-|x|;$ for intermediate $|x| \sim w,$ we assume that $U(x)$ can be well-approximated by a linear potential $U(x) = eEx,$ where $eE$ is a constant. For instance, $U(x) = U \tanh(x/w).$ In the paraxial limit where $k_y$ is small, 
we obtain precisely Eq.~\ref{eq: transmission} as desired. Suppose we start with a localized Gaussian wavefunction at $(x_0, 0)$ on the left side of the PNJ
\begin{equation}
\psi_\text{in} (x,y) = \int \frac{dk_y}{2 \pi} \psi_e(x,k_y) e^{ - a_0^2 k_y^2/2} e^{i k_y y} ,
\end{equation}
where $a_0$ is the width of the wave-packet in coordinate space, and $\psi_e(x,k_y)$ is the electron momentum eigenstate for each $k_y$ on the left side of the PNJ. The transmitted wave packet on the right side of the PNJ is simply 
\begin{equation}
\label{eq: transmitted wave}
\psi_\text{tr}(x,y) = \int \frac{dk_y}{2\pi} \psi_h(x,k_y) e^{-a^2k_y^2/2} e^{i k_y y},
\end{equation}
where $\psi_h(x,k_y)$ is the hole momentum eigenstate for each $k_y$
such that 
\begin{equation}
\label{eq: hole wavefunction}
\psi_h(x,k_y) \propto \exp \left( \frac{i}{\hbar}\int_{x_0}^{-\frac{v_F\hbar |k_y|}{eE}} \phi(\xi) d\xi - \frac{i}{\hbar}\int_{\frac{v_F\hbar |k_y|}{eE}}^{x} \phi(\xi) d\xi \right),
\end{equation} and $a$ is as defined in Eq.~\ref{eq: image size}. For a symmetric PNJ in which $U_\infty = - U_{- \infty},$ $\phi(x)$ is approximately an even function. In this case, the momentum states interfere constructively at $x = -x_0,$ at which point the phase cancels out in Eq.~\ref{eq: hole wavefunction}. This is the point of focusing of fermions on the right side of the PNJ, as shown in Fig.~\ref{fig: trajectories}a. This corresponds to the case with refractive index $n = -1.$

For the case of an asymmetric PNJ, $n = U_{-\infty}/U_{\infty},$ and focusing occurs near $x = n x_0.$ In this case, a cusp is formed near the focal point, as shown in \citep{Cheianov1252} and illustrated in Fig.~\ref{fig: trajectories}b. In both cases of symmetric and asymmetric junctions, the wave packet at the focal point is a Gaussian of momentum eigenstates with width $a$ given by Eq.~\ref{eq: image size}. Thus, the size of the image at the focal point is larger than the original size of the object. The effect of a smooth PNJ is to broaden the image, resulting in imperfect focusing. Eq.~\ref{eq: image size} suggests that the minimal image size is experimentally limited to the ability to fabricate a sharp PNJ. Suppose we have a graphene PNJ with an electrostatic potential of about 20 V, and interface width about 20 nm. Consequently, $eE = 300$ pN, and the minimal image size is approximately 1 nm.

\begin{figure}
\begin{center}
\includegraphics[scale=0.36]{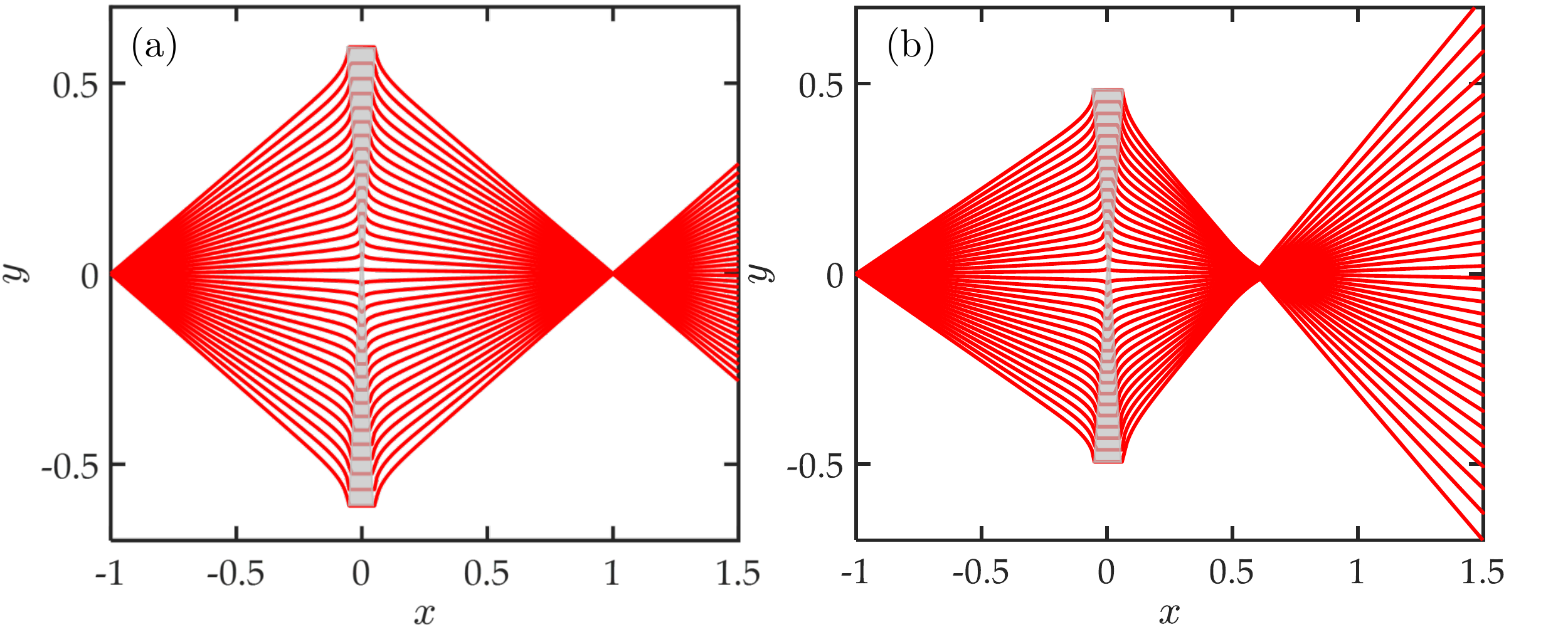}
\caption{Classical trajectories of fermions. We see that in both a symmetric PNJ (a) and an asymmetric PNJ (b), a diverging wave packet on the left side converges to a focus on the right side. The shaded regions indicate tunneling. }
\label{fig: trajectories}
\end{center}\end{figure}

It is interesting to ask whether the sub-wavelength resolution regime (superlensing) can be realized for a smooth potential. To achieve superlensing in this setting, Eq.~\ref{eq: image size} must remain true even when $a \ll \lambda_F,$ the Fermi wavelength. For such to be true,  a necessary condition is $\frac{\pi v_F \hbar}{eE} \ll \lambda^2_F,$ which is satisfied by  increasing the field strength without changing the Fermi wavelength. To see that Eq.~\ref{eq: image size} is true for $a \ll \lambda_F,$ let us consider a symmetric PNJ such that $U_{-\infty} = - U_{\infty}$ with $k_F = |U_{\infty}|/v_F\hbar$ and $\lambda_F = 2 \pi/k_F.$ We follow the same approach as done in \cite{pendry2000negative}. We begin with a localized wavepacket on the left side far away from the junction. For $x \ll 0,$ we can write the wavefunction as a superposition of its Fourier modes as
\begin{align} \label{eq: wavefunction}
\nonumber \psi(x,y) &= \sum_{k_x, |k_y| \leq k_F}\psi(k_x, k_y) e^{ik_xx+ik_yy} \\
&+ \sum_{\kappa, |k_y| > k_F}\psi(k_x, k_y) e^{- \kappa x+ik_yy},
\end{align}
where $k_x = \sqrt{k_F^2-k_y^2}$ and $\kappa = \sqrt{k_y^2-k_F^2}.$ The first part of Eq.~\ref{eq: wavefunction} contains the propagating Fourier modes, while the second part consists of the evanescent waves. As we continue this wavefunction to the right side of the junction, the wavevectors in the $x$-direction change sign, i.e. $k_x \mapsto -k_x$ and $\kappa \mapsto - \kappa$, because we are now in the valence band of graphene. The transmitted component of the wavefunction on the right side far away from the junction with $x \gg 0$ is then
\begin{align} \label{eq: wavefunction on the other side}
\nonumber \psi(x,y) &= \sum_{k_x, |k_y| \leq k_F}t\tilde{\psi}(k_x, k_y) e^{-ik_xx+ik_yy} \\
&+ \sum_{k_x, |k_y| > k_F}t\tilde{\psi}(k_x, k_y) e^{\kappa x+ik_yy},
\end{align}
where the tilde is used to emphasize that the Fourier components on the right side are not necessarily the same as on the left side; and are to be determined by boundary conditions. We see from Eq.~\ref{eq: wavefunction on the other side} that the evanescent components are being amplified on the right side of the PNJ. Thus, at the focal point of the image, all the Fourier components contribute to the image reconstruction. This argument suggests that Eq.~\ref{eq: image size} is true even when the size of the image is smaller than $\lambda_F$, and that lensing below the diffraction limit in a smooth symmetric graphene PNJ is a possibility.

We now consider the question of sub-wavelength resolution in an asymmetric PNJ. In this case, $U_\infty \neq - U_{-\infty}$, and  we need to replace $k_F$ by $k_c = |U_{-\infty}|/v_F \hbar$ on the left side, and $k_F$ by $k_v = |U_{\infty}|/v_F \hbar$ on the right side. Here, we do not get focusing at a single point as shown earlier, but instead get a cusp. However, it is still true that the evanescent waves are also being amplified in this case by an identical argument to the symmetric case. Sub-wavelength resolution is thus possible also in an asymmetric PNJ.

Our calculations suggest that fermionic focusing is a robust feature in smooth graphene PNJs that can potentially operate below the diffraction limit. The presence of finite interface width broadens the image at the focal point, but preserves primary features seen in an abrupt junction. This suggests that graphene is a viable platform to realize fermionic lensing and electronic optics. 

\begin{acknowledgments}
We thank Leonid S. Levitov for suggesting this problem, and for helping us with some of the calculations. We are also grateful to Francisco Guinea and Joaquin Rodriguez-Nieva for critically reading through the manuscript. VTP was supported by the MIT Undergraduate Research Opportunities Program (UROP) for this research. JFK would like to acknowledge financial support from the Singapore A*STAR NSS program.
\end{acknowledgments}

\end{otherlanguage}
\bibliography{lensing_paper}

\end{document}